\documentclass[aps,twocolumn,groupedaddress]{revtex4}
\usepackage[dvips]{color}
\usepackage{graphicx,color}
\begin{document}

\title{Realizing DIII Class Topological Superconductors using $d_{x^2-y^2}$-wave Superconductors}

\author{L. M. Wong, K. T. Law}

\affiliation{Department of Physics, Hong Kong University of Science and Technology, Clear Water Bay, Hong Kong, China}

\begin{abstract}
In this work, we show that a quasi-one-dimensional $d_{x^2-y^2}$-wave superconductor with Rashba spin-orbit coupling is a topological superconductor (TS). This time-reversal invariant DIII class TS supports two topologically protected zero energy Majorana fermions at each end of the system. In contrast to proposals using s-wave superconductors [\onlinecite{STF, SLTD,LSD,Alicea,ORV, BDRv,PL,LSD2, PL11,KMB}] in which a strong magnetic field and the fine tuning of the chemical potential are needed to create Majorana fermions, in our proposal,  the topologically non-trivial regime can be reached in the absence of a magnetic field and in a wide range of chemical potential. Experimental signatures and realizations of the proposed superconducting state are discussed. 
\end{abstract}


\maketitle

\emph{Introduction}---A Majorana fermion is a real fermion which has only half the degrees of freedom of a usual Dirac fermion.  It was first pointed out by Read and Green [\onlinecite{RG}] that  zero energy Majorana fermion modes exist at the vortex cores of a 2D $p_{x}+ip_{y}$ superconductor and these Majorana fermions are non-Abelian particles [\onlinecite{Ivanov}]. Soon after, Kitaev constructed a spinless fermion model and showed that a single Majorana fermion exists at each end of a $p$-wave superconducting wire [\onlinecite{Kitaev00}]. Recently, several groups proposed that effective p-wave superconductors can be realized when an s-wave pairing is induced in systems with spin-orbit coupling [\onlinecite{STF, SLTD,LSD,Alicea,ORV, BDRv,PL,LSD2, PL11,KMB}].

Particularly, the (quasi)-one-dimensional effective p-wave superconductors attracted much attention [\onlinecite{LSD,Alicea,ORV, BDRv,PL,LSD2, PL11,KMB}] due to the fact that Majorana fermion end states can exist in the absence of vortices and the energy separation between the Majorana zero energy mode and other finite energy fermionic modes is relatively large, on the order of the p-wave pairing gap [\onlinecite{PL}].

The existence of Majorana fermions in the above mentioned systems is profound. It is related to the symmetry of the Hamiltonian which decribes the system. According to symmetry classification of Hamiltonians [\onlinecite{SRFL,Kitaev08,TK}], a BdG Hamiltonian with particle-hole symmetry, broken time-reversal  symmetry and broken $SU(2)$ spin rotation symmetry, falls into the D class. In one spatial dimension, a D class Hamiltonian is classified by a $Z_2$ topological number. A system described by a BdG Hamiltonian with a non-trivial $Z_2$ topological number possesses Majorana end states.

To be in the proposed topologically non-trivial regime, it requires a Rashba spin-orbit coupling to break the spin degeneracy, a magnetic field to break the Krammers degeneracy at the Rashba-band crossing point (RCP), fine tuning the chemical potential to the RCP and finally induce an s-wave superconducting pairing at the Fermi energy. However, tuning the chemical potential to the RCP, which is near the electronic band bottom, reduces the electron density severely and electrons can be easily localized by disorder in this regime. The strong magnetic field required can also suppress superconductivity. A schematic picture of this proposal is shown in Fig.2a.

A DIII class Hamiltonian respects both time-reversal and particle-hole symmetry and breaks $SU(2)$ spin-rotation symmetry [\onlinecite{SRFL,Kitaev08,TK}].  In this Letter,  we point out that a  DIII class TS can be realized when electrons in  a quasi-one-dimensional wire with spin-orbit coupling acquire a  $d_{x^2-y^2}$-wave pairing. In the topologically non-trivial regime, two zero energy Majorana fermion modes appear at each end of the wire. In our proposal, Majorana fermions can be created in the absence of an external magnetic field and in a wide range of chemical potential, e.g. there is no need to tune the chemical potential to the RCP. In the presence of an external magnetic field, the system is in the D class and a single Majorana end state appears at each end of the wire. Experimental signatures and realizations of this DIII class TS will be discussed at the end.

\emph{Single-channel model}--- Before studying the more realistic quasi-one-dimensional quantum wires,  in this section,  we first consider a strictly one-dimensional DIII class Hamiltonian which supports double Majorana end states in the absence of an external magnetic field. We construct the following model
\begin{equation}
\begin{array}{l}
H_{1D}= H_{t} + H_{SO}+H_{SC} +H_{Z} \\
H_{t}= \sum_{j,\alpha} -\frac{t}{2}(\psi^{\dagger}_{j+1,\alpha}\psi_{j \alpha}+h.c.)-\mu \psi^{\dagger}_{j,\alpha}\psi_{j \alpha}\\
H_{SO}=\sum_{j,\alpha,\beta}-\frac{i}{2} \alpha_{R} \psi^{\dagger}_{j+1,\alpha} (\sigma_{y})_{\alpha,\beta} \psi_{j,\beta}+h.c.\\
H_{SC}=\sum_{j} \frac{1}{2}\Delta_{0} (\psi^{\dagger}_{j+1,\uparrow}\psi^{\dagger}_{j,\downarrow}-\psi^{\dagger}_{j+1,\downarrow}\psi^{\dagger}_{j,\uparrow})+h.c. \\
H_{Z}=\sum_{j} V_{z}(\psi^{\dagger}_{j\uparrow}\psi_{j\uparrow}-\psi^{\dagger}_{j\downarrow} \psi_{j\downarrow}),
\end{array} \label{H1d}
\end{equation}
where $\psi_{j}$ is a fermion operator at site $j$, $\alpha$ and $\beta$ are the spin indices, $t$ is the hopping amplitude, $\alpha_{R}$ is the spin-orbit coupling strength, $\Delta_{0}$ is the superconducting pairing amplitude, and $\sigma_y$ is a Pauli spin matrix. $V_z$ denotes the strength of the Zeeman term.  Without the pairing terms, the above model describes a wire with spin-orbit coupling. If an s-wave (on-site superconducting) pairing is induced on the wire, as it is done in Ref.[\onlinecite{LSD, Alicea,ORV, BDRv,PL,LSD2,PL11,KMB}], no Majorana fermions can be created without breaking time-reversal symmetry. In the following, we show that our model supports double Majorana end states in the presence of time-reversal symmetry.

The energy spectrum of $H_{1D}$ with $V_z=0$ is shown in Fig.1a. Due to Krammers degeneracy, every energy level in Fig.1a is doubly degenerate. It is evident from the energy spectrum that zero energy modes exist when the chemical potential satisfies $|\mu| < |\alpha_{R}|$. The sum of the two ground-state wavefunctions is shown in Fig.1b to comfirm that the zero energy modes are end states. In other words, there are two Majorana fermions at each end of the wire. In the topologically trivial regime where $|\mu| > |\alpha_{R}|$, the ground state wavefunctions are predominantly in the bulk as shown in Fig.1c.

\begin{figure}
\includegraphics[width=3in]{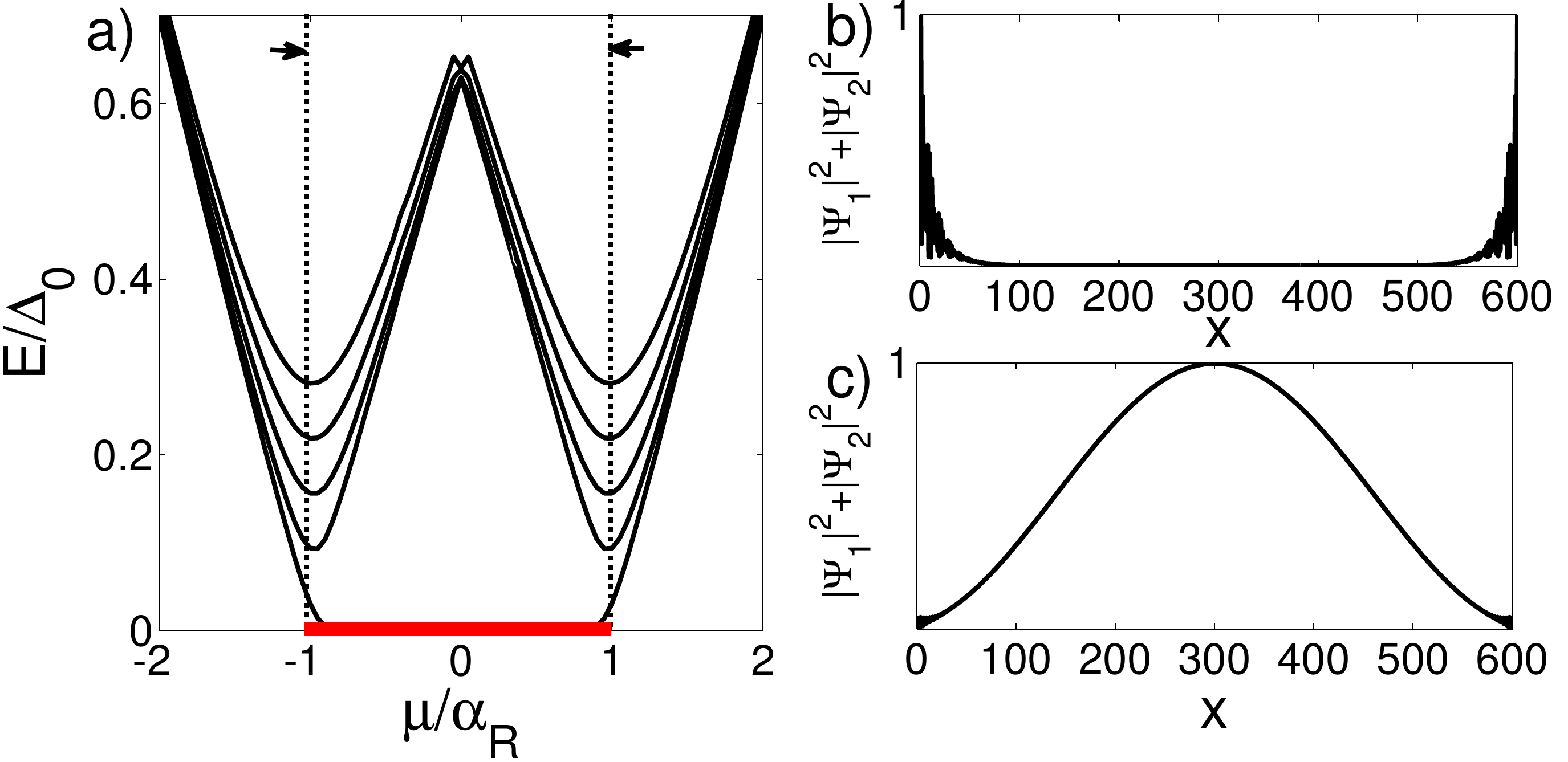}
\caption{\label{Fig1} a) Excitation energy versus chemical potential. The parameters of $H_{1D}$ are: $L=600$, $t=12$, $\Delta_0=2$ and $\alpha_{R}=4$, where $L$ is the number of sites. Zero-energy modes exist in the region bounded by the vertical dotted lines where $|\mu| < |\alpha_R|$. All the bands shown in the figure are doubly degenerate due to time-reversal symmetry. b) $|\Psi_{1}|^2+|\Psi_{2}|^2$ versus $x$ where $x$ is the site label. The ground-state wavefunctions are localized at the edge in the topologically non-trivial regime $\mu=\alpha_{R}/2$. c) In the topologically trivial regime, e.g. $\mu=1.5 \alpha_{R}$, the ground-state wavefunctions are predominantly in the bulk.}
\end{figure}

To understand how the DIII topological superconducting state is achieved in our model, we note that a general criteria for realizing a one-dimensional DIII class TS is to have an odd number of negative pairing amplitude at the Fermi points with Fermi momentum between $0$ and $\pi$ [\onlinecite{QHZ}]. We show that this is indeed the case for $H_{1D}$.

In the momentum space, Hamiltonian $H_{1D}$ can be written as
\begin{equation}
\begin{array}{ll}
H_{1D}(k)=   \sum_{k} & \Psi_{k }^{\dagger} [-(t\cos k + \mu) \sigma_{0} + \alpha_R \sin k \sigma_{y}] \Psi_{k} + \\ &  \Delta_{0} \cos k \psi_{k \uparrow}^{\dagger} \psi_{-k \downarrow}^{\dagger} + h.c.
\end{array}
\end{equation}
where $\Psi_{k}^{\dagger}=(\psi_{k \uparrow}^{\dagger}, \psi_{k \downarrow}^{\dagger})$. The Hamiltonian has spectrum $E(k)=\pm \sqrt{((-t\cos k - \mu) \pm \alpha_{R} \sin k)^2+ (\Delta_0 \cos k)^2}$ and it is generally gapped unless $|\mu|=\alpha_R$ at which points topological phase transitions take place. In the basis which diagonalize the Rashba term, the Hamiltonian  can be written as 
\begin{equation}
\begin{array}{ll}
\tilde{H}_{1D}(k)=  \sum_{k, a=\pm} &  [-(t\cos k + \mu) +a |\alpha_R \sin k| ] \tilde{\psi}_{k a}^{\dagger}  \tilde{\psi}_{k a}  + \\ & \text{sgn}(k) \Delta_{0} \cos k \tilde{\psi}_{k a}^{\dagger}  \tilde{\psi}_{-k a}^{\dagger} + h.c. ,
\end{array}
\end{equation}
where $\tilde{\psi}_{k a}$ denotes a fermion in the new band basis. When $|\mu| < |\alpha_{R}|$, there are two Fermi points $k_{1}, k_{2}$ with  $0 < k_{1}, k_{2} < \pi$. In this regime, it can be shown that one and only one of the pairing amplitudes of the two bands at the Fermi level $\Delta_0 \cos k_{1} $ and $\Delta_0 \cos k_{2} $ is always negative. Therefore, the superconductor is in the topologically non-trivial regime. A schemetic picture is shown in Fig.2b. When $|\mu|>|\alpha_{R}|$, we have $\cos k_{1} \cos k_{2} >0$ and the pairing amplitudes at the two Fermi points have the same sign and the system is in the topologically trivial regime.

\begin{figure}
\includegraphics[width=3in]{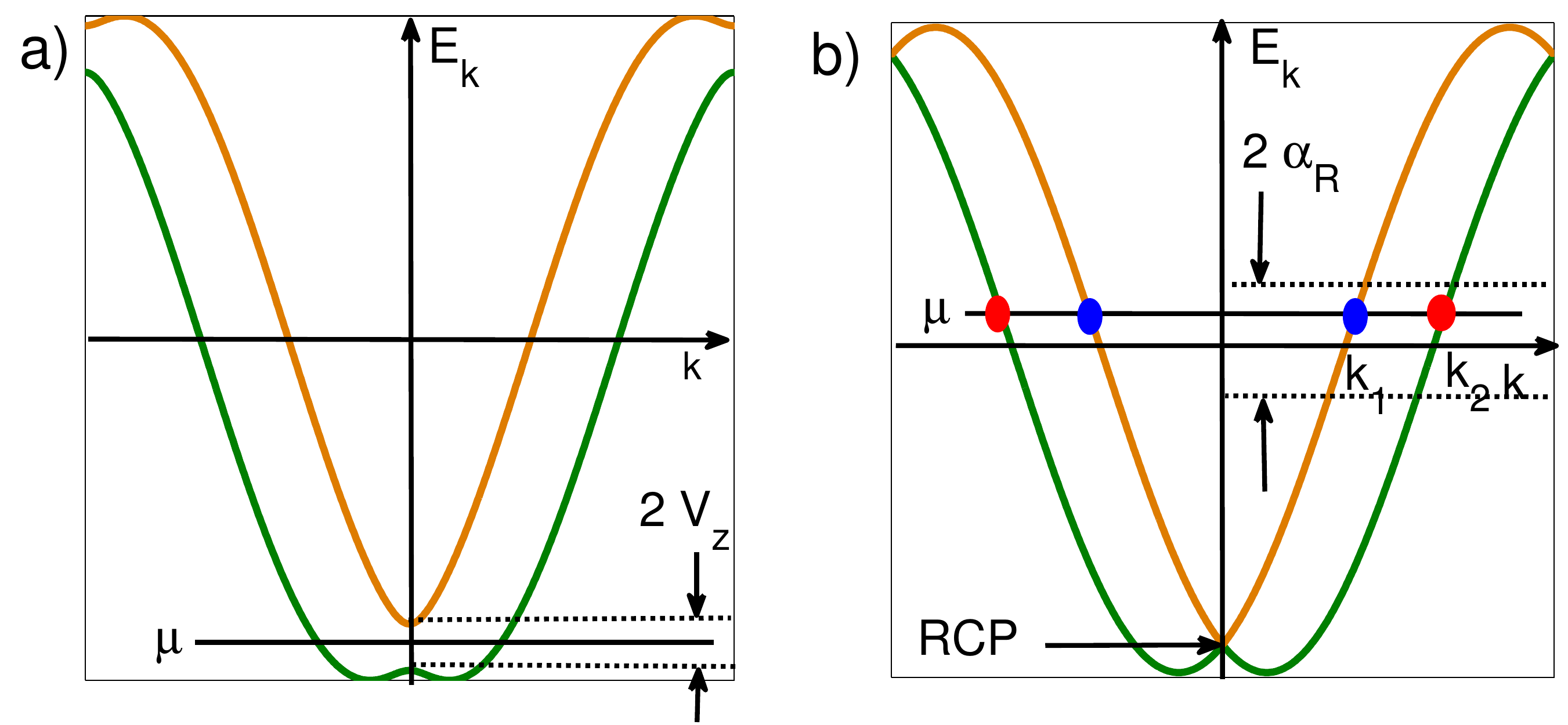}
\caption{\label{Fig2} a) The energy versus momentum of a typical one-dimensional Rashba band with a finite Zeeman term. In s-wave-pairing proposals, a strong magnetic field is needed to break the Krammers degenercy at the RCP. The chemical potential needs to be tuned to  the topologically non-trivial regime which is bounded by the horizontal dashed lines near the electronic band bottom. Electron density in this regime is low. b) In our one-dimensional model with $V_z=0$, topologically non-trivial regime is located at the middle of the band which is bounded by the dashed lines. In this regime, the pairing amplitudes  $\Delta_0 \cos k_1$ and $ \Delta_0 \cos k_2$ at the two Fermi points, $k_1$ and $k_2$, have opposite signs.   }
\end{figure}

In short, in order to reach the topologically non-trivial regime, we need to break the spin degeneracy by the Rashba term and induce a $k$-dependent pairing such that there can be an odd number of negative pairing amplitudes for positive $k$ at the Fermi energy. It is important to note that there is no need to tune the chemical potential to the RCP which is near the band bottom.  If the induced pairing is s-wave [\onlinecite{LSD, Alicea,ORV, BDRv,PL,LSD2,PL11,KMB}], the topologically non-trivial phase is not accessible.

The topologically non-trivial state can be further verified by calculating the topological invariant of $H_{1D}(k)$ [\onlinecite{QHZ}]. The topological invariant can be written as
\begin{equation}
N_{DIII}=\frac{\text{Pf}[Tq(k=\pi)]}{\text{Pf}[Tq(k=0)]} \text{exp}\{-\frac{1}{2} \int_{0}^{\pi} dk \text{Tr}[q^{\dagger}(k)\partial_{k}q(k)]\},   \label{DIIITI}
\end{equation}
where $\text{Pf}$ denotes the Pfaffian,  $T=i\sigma_{y}$ is the time-reversal operator, and $q(k)=\frac{1}{2} [e^{i\theta_{-}(k)}(\sigma_{0}-\sigma_{y})  + e^{i\theta_{+}(k)}(\sigma_{0}+\sigma_{y}) ]$ which is an off-diagonal block of the flat-band Hamiltonian [\onlinecite{QHZ,SR}] derived from $H_{1D}(k)$. Here, $e^{i\theta_{\pm}}=\frac{-t\cos(k)-\mu\pm \alpha_{R}\sin(k)+i \Delta_{0}\cos(k)}{\sqrt{[-t\cos(k)-\mu \pm \alpha_{R}\sin(k)]^2+ [\Delta_{0}\cos(k)]^2}}$. From Eq.\ref{DIIITI}, the topological invariant number can be found to be trivial ($N_{DIII}=1$) when $|\mu| > \alpha_{R} $ and non-trivial ($N_{DIII}=-1$) when $|\mu| < \alpha_{R}$.

\emph{Single-channel model with finite $V_z$}---When $V_z$ is finite, the energy spectrum of $H_{1D}(k)$ becomes $E(k)=\pm\sqrt{F(k) \pm 2\sqrt{G(k)}}$,
where $F(k)=(t\cos k +\mu)^2+\alpha^{2}_{R}\sin^2 k+\Delta^2_{0}\cos^2 k+ V^2_{z}$ and $G(k)=(t\cos k +\mu)^2 V^{2}_{z}+(t\cos k +\mu)^2 \alpha^{2}_{R}\sin^{2}k +V^{2}_{z}\Delta^2_{0}\cos^2 k$. From the energy spectrum, we note that the energy gap closes when $(\mu \pm t)^2=V^2_z-\Delta^2_{0}$ and $|\mu|=\sqrt{V^2_z+\alpha^2_{R} }$.

Moreover,  the $V_z$ term breaks time-reversal symmetry and changes the Hamiltonian from DIII class to D class. It is known that a 1D Hamiltonian in D class may support single Majorana end states [\onlinecite{Kitaev00, SRFL}]. The energy eigenvalues of $H_{1D}$ versus the chemical potential are shown in Fig.3. In Fig.3, the region between $c_{1}$ and $c_2$ allows double Majorana fermions where $c_{1}$ and $c_2$ are points with $\mu=\mp|\sqrt{V^2_z-\Delta^2_{0}}- t|$ respectively [\onlinecite{DM}]. In the regions between $a_{1}$ and $c_1$, $c_2$ and $a_2$, single Majorana fermion end states emerge. Here, $a_{1}$ and $a_2$ indicate points with $\mu =\mp| \sqrt{V^2_z-\Delta^2_{0}} + t|$ respectively. When $|\mu| > |\sqrt{V^2_z-\Delta^2_{0}} + t|$, the Hamiltonian is topologically trivial and the Majorana end states disappear. At points $b_1$ and $b_2$ where $\mu=\mp \sqrt{V^2_z+\alpha^2_{R} } $, the energy gap closes but there are no topological phase transitions at these points.

\begin{figure}
\includegraphics[width=3in]{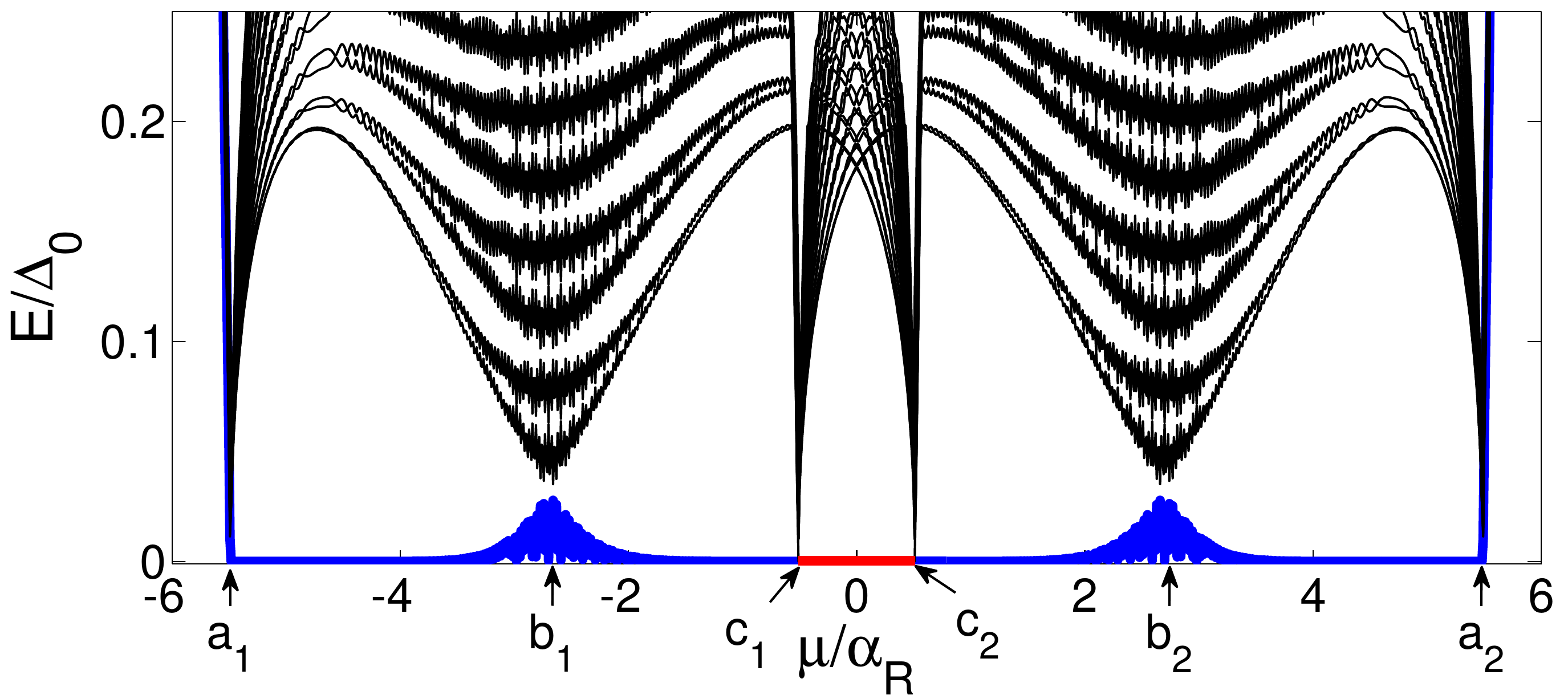}
\caption{\label{Fig3} Excitation energy as a function of chemical potential. The parameters of $H_{1D}$ are: $t=12$, $\Delta_0=1$, $\alpha_R=4$ and $V_z=10$. Points $a_{i}$, $b_{i}$ and $c_i$ (with i=1,2) denote gap closing points with chemical potential $\mu=\mp|\sqrt{V^2_z-\Delta^2_{0}} + t|$,  $\mu=\mp |\sqrt{V^2_z+\alpha^2_R}|$ and $\mu=\mp |\sqrt{V^{2}_{z}-\Delta^{2}_{0}}-t|$  respectively. Double Majorana fermion end states exist when $\mu$ is between $c_1$ and $c_2$. Single Majorana fermion end states exist when $\mu$ is in the regions between $a_1$ and $c_1$, $c_2$ and $a_2$.}
\end{figure}

The even and odd number of Majorana end states in Fig.3 can be verified by calculating the $Z_2$ Majorana number $\mathcal{M}$ of $H_{1D}(k)$. Following Refs [\onlinecite{Kitaev00,LSD2}] the Majorana fermion number can be defined as
\begin{equation}
\mathcal{M}=\text{sgn}[\text{Pf}B(0)]\text{sgn}[\text{Pf}B(\pi)]=\pm 1.   \label{MN}
\end{equation}
The matrix $B(k)$ is defined as $B(k)=H_{1D}(k)(\sigma_{x}\otimes \sigma_{0})$. $B(k)$ is anti-symmetric and its Pfaffian is well-defined. $\mathcal{M} = \pm 1$ indicates the even and odd number of Majorana fermions at one end of the wire respectively. In terms of the parameters of the Hamiltonian, the Majorana number can be written as
\begin{equation}
\text{sgn}\{[(t+\mu)^2-(V^2_{z}-\Delta^{2}_{0})][(-t+\mu)^2-(V^{2}_{z}-\Delta^{2}_{0})]\}. \label{Mnumber}
\end{equation}
The Majorana numbers calculated according to Eq.\ref{Mnumber} are consistent with the results shown in Fig.3. The Majorana number is $\mathcal{M}=-1$ when $|\sqrt{V^2_z-\Delta^2_{0}}- t|<|\mu|<|\sqrt{V^2_z-\Delta^2_{0}} +t|$ and $\mathcal{M}=1$ otherwise.

\emph{Multi-channel case}---In realistic situations, multiple transverse sub-bands of a wire are occupied and it is important to show that Majorana fermions exist in this situation. In this section, we show that Majorana fermions exist in quasi-one-dimensional wires. Importantly, the quasi-one-dimensional model can be realized by inducing $d_{x^2-y^2}$-wave superconductivity on a wire with strong spin-orbit coupling.

In the quasi-one-dimensional case, the Hamiltonian can be written as:
\begin{equation}
\begin{array}{ll}
H_{q1D} = & H_{t} + H_{SO}+H_{SC} +H_{Z}, \\
H_{t}= & \sum_{\mathbf{R},\mathbf{d}, \alpha} -\frac{1}{2}t(\psi^{\dagger}_{\mathbf{R+d},\alpha}\psi_{\mathbf{R} \alpha}+h.c.)-\mu \psi^{\dagger}_{\mathbf{R},\alpha}\psi_{\mathbf{R} \alpha}\\
H_{SO}= & \sum_{\mathbf{R,d},\alpha,\beta}-\frac{i}{2} \alpha_{R} \psi^{\dagger}_{\mathbf{R+d},\alpha} \hat{\mathbf{z}}\cdot(\vec{\sigma}_{\alpha \beta}\times \mathbf{d})\psi_{\mathbf{R},\beta}+h.c.\\
H_{SC}= & \sum_{\mathbf{R}}\frac{1}{2}[ \Delta_{0} (\psi^{\dagger}_{\mathbf{R+d_{x}},\uparrow}\psi^{\dagger}_{\mathbf{R},\downarrow}-\psi^{\dagger}_{\mathbf{R+d_{x}},\downarrow}\psi^{\dagger}_{\mathbf{R},\uparrow})- \\ 
&\Delta_{0}(\psi^{\dagger}_{\mathbf{R+d_{y}},\uparrow}\psi^{\dagger}_{\mathbf{R},\downarrow}-\psi^{\dagger}_{\mathbf{R+d_{y}},\downarrow}\psi^{\dagger}_{\mathbf{R},\uparrow}) +h.c. ]\\
H_{Z}= & \sum_{\mathbf{R}} V_{z}(\psi^{\dagger}_{\mathbf{R}\uparrow}\psi_{\mathbf{R}\uparrow}-\psi^{\dagger}_{\mathbf{R}\downarrow} \psi_{\mathbf{R}\downarrow}).
\end{array}  \label{Hq1D}
\end{equation}

Here, $\mathbf{R}$ denotes the lattice sites, $\mathbf{d}$ denotes the two unit vectors $\mathbf{d_{x}}$ and $\mathbf{d_y}$ which connects the nearest neighbor sites in the $x$ and $y$ directions respectively. This model is the same as the tight-binding model in Ref[\onlinecite{PL11}] except for the superconducting pairing terms. The pairing terms in $H_{q1D}$ can be written as $\Delta_{0}[ \cos(k_{x})- \cos(k_{y})]$ in the momentum space. Therefore, $H_{q1D}$ describes a quantum wire with spin-orbit coupling and a $d_{x^2-y^2}$-wave superconducting pairing.  A schematic picture of the experimental setup is shown in Fig.4.

\begin{figure}
\includegraphics[width=3in]{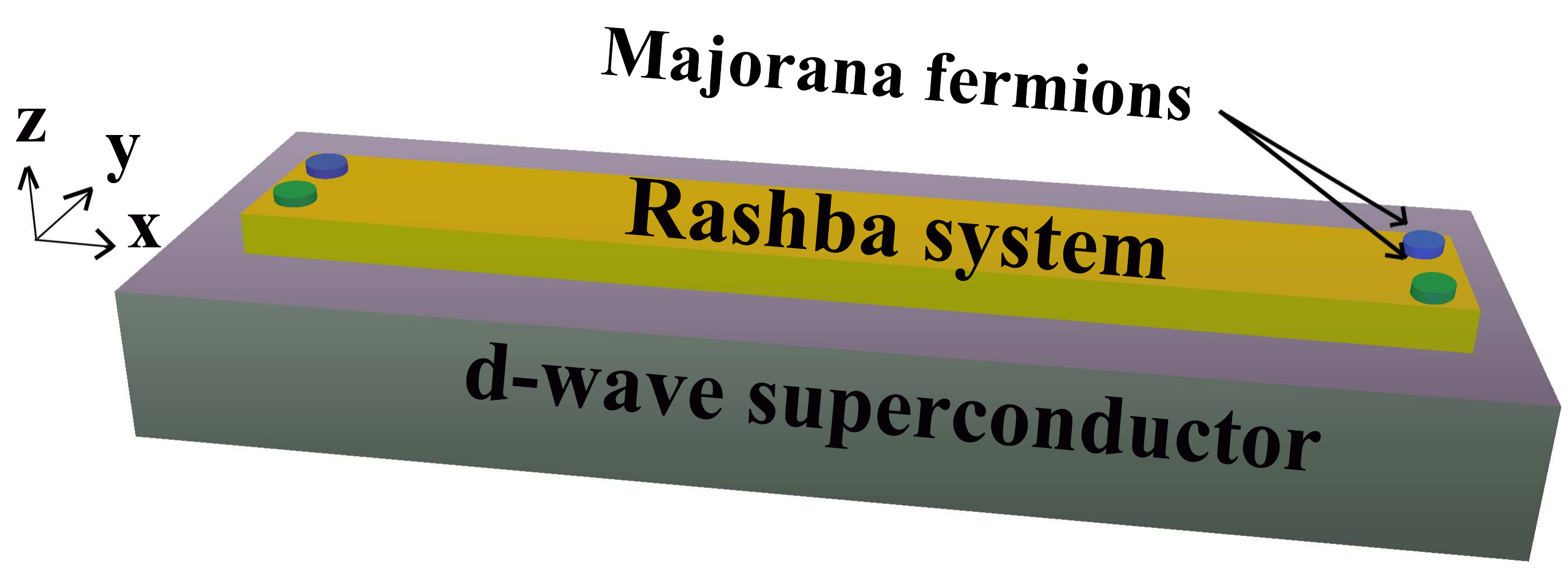}
\caption{\label{Fig4} A wire with strong spin-orbit coupling in proximity to a $d_{x^2-y^2}$-wave superconductor.  Double Majorana end states may appear in the absence of an external magnetic field. }
\end{figure}

The energy spectrum of the Hamiltonian with $V_z =0$ is shown in Fig.5a. The length of the wire is chosen to be much larger than the superconducting coherence length $L \gg t / \Delta_0 $ and the width is comparable to the coherence length $W \approx t/\Delta_0$. In sharp contrast to the s-wave-pairing proposals in which the topologically non-trivial regime can be reached only when the chemical potential is near the RCP, in our proposal, zero energy Majorana modes live all over the full band even when the Fermi level is far away from the band bottom. It is also evident from Fig.5a that the zero energy Majorana modes are separated from other fermionic states by a large energy gap which is of order $\Delta_0$. Due to time-reversal symmetry, all energy levels in Fig.5a is doubly degenerate. Therefore, in the topologically non-trivial regime, there are two Majorana fermions at each end of the wire. It can be shown that these double Majorana end states are robust to disorder.

\begin{figure}
\includegraphics[width=3in]{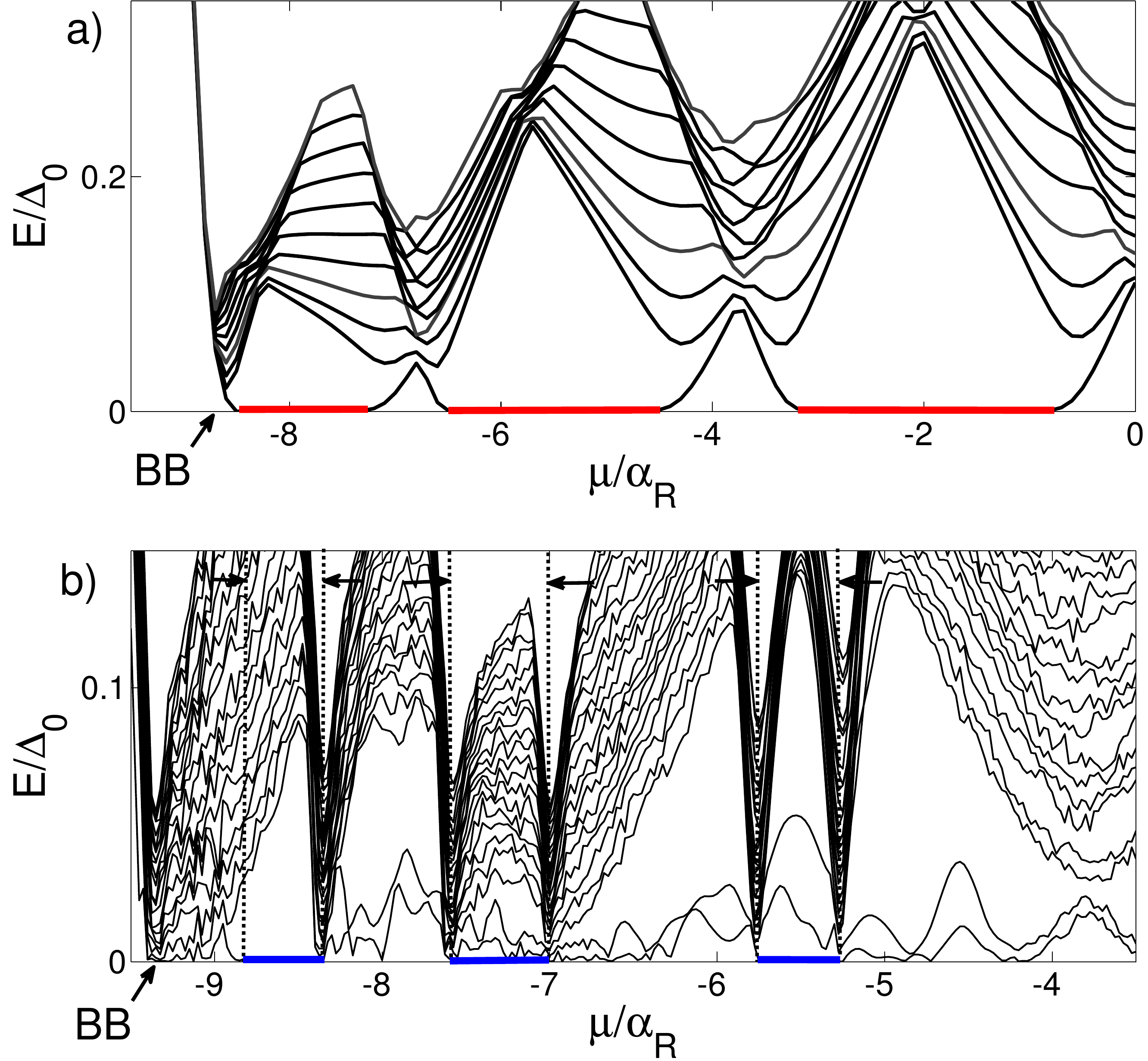}
\caption{\label{Fig5}  a) Excitation energy versus chemical potential. The parameters in $H_{q1D}$ are $V_z=0$, $t=18$, $\Delta_0=2$ and $\alpha_R=4$. The width of the wire is $W=6$ and the length is $L=600$. All the states shown are doubly degenerate. Zero energy Majorana modes, denoted by the red lines, can appear in a wide range of chemical potential. The normal state electronic band bottom is denoted as BB b) Excitation energy versus chemical potential in the presence of random potential. The parameters in $H_{q1D}$ is the same as in a) except $V_z=3$. The topologically non-trivial regimes with an odd number of zero energy modes are bounded by dotted lines. }
\end{figure}

In the presence of the $V_z$ term, time-reversal symmetry is broken and the Hamiltonian is in D class. To break some incidental degeneracy, a random on-site potential $V$ with strength $\sqrt{\langle V^2 \rangle -\langle V \rangle ^2}=\Delta_0 /2$ is introduced into the system. The resulting energy spectrum of the model with $V_z=3$ is shown in Fig.5b. Without time-reversal symmetry, the spectrum is no longer doubly degenerate. From Fig.5b, it is evident that zero energy modes which represent single Majorana end states exist. As expected, the single Majorana modes are more robust when the chemical potential is near the band bottom. Majorana fermions exist when the chemical potenital is further away from the band bottom, however, the mini-gap, which is the energy gap between the Majorana modes and other finite energy fermionic modes, is much smaller as in the s-wave pairing case [\onlinecite{PL12}].

It is shown previously that a single Majorana fermion induces a zero bias conductance peak of $G=2\frac{e^2}{h}$ in Andreev reflection experiments [\onlinecite{LLN, WADB}] when a normal metal lead couples to a Majorana end state. On the other hand, the double Majorana end states in DIII class TS induce a zero bias conductance peak of $G=4\frac{e^2}{h}$ [\onlinecite{FHAB,LiuLN}] instead. Therefore, the two topological superconducting phases can be distinguished from each other and from trivial superconducting states through Andreev reflection experiments.

\emph{Discussion}--- A few important comments follow. First, the DIII class topological superconductor proposed in this work is truly different from the time-reversal invariant superconducting state obtained by inducing an s-wave or a d-wave pairing on the surface of a topological insulator (TI) [\onlinecite{FK,LTYSN,CF}]. The TI surface state is described by a single Dirac cone (in the simplest case). With only a single species of fermion at the Fermi level, one cannot create a superconducting wire with double Majorana end states even though single Majorana fermions can be created in the presence of an external magnetic field. Sato et al. studied a d-wave superconductor with Rashba terms in the presence of an external magnetic field but no DIII class TS phase is reported [\onlinecite{SF}].

Second, it is pointed out recently that single Majorana fermion end states can be created at arbitary chemical potential if  a magnetic field is applied along the wire in the s-wave pairing case [\onlinecite{CF,PL12}]. However, the magnetic field $\vec{B}$ required to reach the topologically non-trivial state is still strong comparing to the induced pairing gap $\Delta_S$, namely, $\mu_0 |\vec{B}| > \Delta_S$ where $\mu_0$ is the effective magnetic moment of electrons. 

Third, for simplicity, we assumed that the wire is aligned along the x-direction. Tilting the wire with respect to the x-axis slightly has no major effect on the Majorana end states since this kind of perturbation does not change the symmtry class of the system.

Fourth, one possible realization of $H_{q1D}$ is by inducing $d_{x^2-y^2}$-wave superconducting pairing on the Au (111) surface state. It has been shown that the Au (111) surface has a  Rashba band with Rashba energy of about 60meV [\onlinecite{HEB}]. The proximity gap induced on Au by LSCO can reach 10meV [\onlinecite{YAKKM}]. These large energy scales make Au on LSCO a promising candidate of realizing topological superconducting states. However, the induced proximity pairing shown in the recent experiment may not be d-wave due to the presence of strong disorder in the Au layer. 

Another candidate material of a DIII class TS is a layered heavy fermion superconductor $\text{CeCoIn}_{5}$. Bulk $\text{CeCoIn}_{5}$ is a $d_{x^2-y^2}$-wave superconductor. Recently, superconducting thin films of $\text{CeCoIn}_{5}$ with only several atomic layers thick can be fabricated [\onlinecite{matsuda1}]. Suppose we have a thin film of $\text{CeCoIn}_{5}$,  due to the strong spin-orbit coupling and the broken of inversion symmetry on the surface, the top layer of $\text{CeCoIn}_{5}$ acquires a Rashba term [\onlinecite{Sigrist}]. Therefore, the top layer of $\text{CeCoIn}_{5}$ can be described by $H_{q1D}$. In a separate work, we show in detail that a thin film of $\text{CeCoIn}_{5}$ is a DIII class TS.

\emph{Conclusion}--- We show that a quasi-one-dimensional $d_{x^2-y^2}$-wave superconductor with Rashba spin-orbit coupling terms is a DIII class TS which supports double Majorana end states. In the presence of a magnetic field, single Majorana end states appear. These two topological superconducting states can be probed using Andreev reflection experiments. We suggest that Au wires on LSCO and thin films of $\text{CeCoIn}_{5}$ are candidate materials of this new topological superconducting state.

\emph{Acknowledgments}--- We thank C.H. Chung,  C.Y. Hou, J. Liu, T.K. Ng, B. Normand and A. Potter for insightful discussions. We thank Y. Matsuda for bringing the $\text{CeCoIn}_{5}$ system to our attention. KTL thanks Patrick Lee for being a source of inspiration. The authors are supported by HKRGC through DAG12SC01 and HKUST3/CRF09.


\begin{thebibliography}{99}
\bibitem{DM} Due to the fact that $H_{1D}$ is real, one can construct a $Z_n$ topological invarant and show that the double Majorana zero modes are robust against disorder even when $V_z$ is finite. However, this is truly only for the strictly one-dimensional case.
\end{thebibliography}
\end{document}